\documentclass[11pt,fleqn]{article}
\usepackage[dvips]{graphicx} 
\begin{document}
\title{{\LARGE \bf  Self-Referential Noise and the Synthesis of Three-Dimensional Space}
 \\ \vspace{6mm}
{\normalsize \bf - 1998 Heraclitean Process System  Report -}}  
\author{{Reginald T. Cahill  and     Christopher M. Klinger}\\
  {Department of Physics, Flinders University
\thanks{E-mail: Reg.Cahill@flinders.edu.au, Chris.Klinger@flinders.edu.au}}\\ { GPO Box
2100, Adelaide 5001, Australia }}
\date{ }

\maketitle

\vskip 0.6in
\begin{center}{\bf Abstract}\end{center}
Generalising results from G\"{o}del and  Chaitin in mathematics suggests that  self-referential
systems contain intrinsic randomness.  We argue that this is relevant to modelling the universe and
show how three-dimensional space may arise from a non-geometric order-disorder model driven by
self-referential noise.

\vspace{10mm}


\newpage

\section  {\bf  Introduction\label{section:Introduction}} 

General relativity begins the modelling of reality by assuming differentiable manifolds and
dynamical equations for a $3\oplus1$ metric spacetime.  Clearly this amounts to a high level
phenomenology which must be  accompanied by various meta-rules for interpretation and application. 
The same situation also occurs in the quantum theory.  But how are we to arrive at an understanding 
of the origin and  logical necessity for the form of these laws and their interpretational
meta-rules?  We argue  that the understanding, and therefore also the unification, of these almost
fundamental theories  actually involves an appreciation of the special features  of  {\it
end-game} modelling.  In essence an {\it end-game} theory is one which involves the unique
situation  in which we attempt to  model reality without {\it a priori} notions.  Present day
theoretical physics  is very much based upon the notion of {\it objects} and their rules, which
amounts to {\it numbers} (of objects) and thus arithmetic, and the generalisation to sets and
abstract mathematics. From these rules of objects we also arrived at traditional {\it object-based
logic}. Axiom-based modelling always assumes some starting set of {\it objects} and their
{\it rules}.  This has the apparent defect of requiring   an infinite regress of a hierarchy of
models; each one yielding a higher level as an emergent phenomenon.  Here  this profound problem of an
infinite regress of nested {\it objects} and their  {\it rules}   is overcome by proposing that a
fundamental modelling of reality  must invoke self-organised criticality (SOC) to hide the start-up
axioms via {\it universality}.  Further we also generalise the results of G\"{o}del and Chaitin in
mathematics to argue that a self-referential system, such as the universe, must involve intrinsic
randomness, which we name Self-Referential Noise (SRN). 
 Our analysis of {\it end-game} modelling then 
results in a sub-quantum non-geometric order-disorder  process model  driven by SRN
from which, the evidence suggests,  emerges a fractal 3-space; the fractal character being necessary
to achieve {\it universality}.  Self-consistency requires that at higher levels {\it objects} and
their {\it rules} emerge via an {\it objectification} process, together with  the quantum
phenomena. We call this approach a Heraclitean Process System (HPS) after Heraclitus  of
Ephesus (6 BCE) who, in western science,  first emphasised the importance of process over
object.  We suggest that the SRN provides an important extension to the pregeometric class of
modellings of reality\cite{Wheeler80,Gibbs,Finkelstein}.

\section  { Self-Referential Noise\label{section:Self-Referential-Noise}}  

Our proposed solution to the end-game problem is thus to avoid the notion of {\it objects} and their
{\it rules} as fundamental; for these categories are only appropriate for higher level
phenomenological modelling.    However the problem is then not merely to construct some  model of 
 reality but to do so using the, in principle, inappropriate {\it language} of objects which
themselves are high level emergent phenomena; i.e. we must develop a technique to extend object-based
logic and mathematics  beyond their proper domains. We propose to achieve this by exploiting the
concept of {\it universality} in self-organising criticality (SOC).  SOC describes the ability of
many systems (the first studied being the sand-pile\cite{SOC}) to self-organise in such a way that
the system itself moves towards a state characterised by a fractal description, i.e. in which there
is no fundamental scale and phenomena of all scales appear (avalanches in the case of the sand-pile
model\cite{SOC}).  This fractal or universality  property had been much investigated 
in non-SOC systems in which the universality only appeared at critical points that could be
reached by suitably adjusting external parameters such as temperature and pressure. The term
{\it universality} indicating that the behaviour of the system at such a critical point was not
uniquely characteristic  of individual systems, but that many systems, in the same {\it
universality class}, showed the same behaviour; at criticality the individuality of the
system is suppressed. SOC systems display the novel behaviour of always evolving towards such
criticality without the need for fine tuning of external parameters. Smolin\cite{Smolin}
has discussed the possible relevance of SOC  to cosmology. 
 
 The notion that a system is self-supporting or {\it bootstrapped} has always been weakened so that
some set of axioms is invoked as a start-up part of the system.  In HPS we require that the start-up
axioms be suppressed by requiring a SOC system for which many other modellings belong to the same
universality class. In this way we attempt to  achieve  an {\it axiom-less}  model. 
Hence in the HPS modelling of reality we apply object-based logic and mathematics
to  an individual realisation of a, hopefully,  SOC relational process, but then confirm and
extract the {\it universal} emergent behaviour which will  be independent of the
realisation used.   To be consistent  any HPS must not only display SOC but at some high
level it must also display an emergent  objectification process which will be
accompanied by its  object-based logic. The key idea is that a truly {\it bootstrapped}
model of reality must self-consistently bootstrap {\it logic} itself, as well as the {\it
laws of physics}.  Further, only by constraining our modelling to such a complete bootstrap do
we believe  we can arrive at complete comprehension of the nature of reality. 
 
Together with
the SOC process we must also take account of the powerful notion that by definition a universe
is self-contained, and this also has profound implications for our {\it ab initio} bootstrap
modelling. This means, as we will argue, that the universe is necessarily self-referential, and
that this suggests that we must take cognisance of a  fundamental and irremovable non-local
randomness (SRN). We propose
here that SRN  is a fundamental process that has been ignored until now in model building in
physics and in the general comprehension of reality. Again our SOC proposal is to be invoked in order
that the SRN not be understood as a `thing', but as a realisation-independent characterisation of the
self-referencing.  Nevertheless we believe that already there are  several indicators of
the special consequences of this SRN, but not understood as such in present day physics. 
  
The construction of a viable HPS can only be achieved at present by inspired guessing based in
part upon the lessons of Quantum Field Theory (QFT) and also the consequences of the
self-referential  process in mathematics where the precision of analysis by G\"{o}del and
others has lead to  definitive conclusions.  In QFT there are  several features that have
suggested to us a deeper processing; first, that all practical non-perturbative computations
are done most easily and efficaciously in the Euclidean rather than Minkowski metric.  In
this Euclidean metric it is more natural to think of the QFT as an ensemble average of a zero
temperature statistical system.  Furthermore, these zero-temperature functional-integral
ensemble averages can be obtained as the ensemble averages of Wiener processes  via the {\it
stochastic quantisation} construction\cite{Stochastic}; a construction which has lacked until
now an  interpretation.    However the key clue is that  the  {\it stochastic
quantisation} (where the term {\it quantisation}  now appears inappropriate) invokes a random noise
process which here we identify with the SRN. 

That the modelling of  a self-referential system contain SRN we believe follows plausibly as a
generalisation from the work of Chaitin\cite{Chaitin} 
 who, extending the work of G\"{o}del and  Turing,  
 showed that the arithmetic system is sufficiently complex that the self-referential
capability of arithmetic results in randomness and unpredictability,  and 
means that in some manner arithmetic should be thought of in a thermodynamic  sense.  Patton and
Wheeler\cite{PW} conjectured some time ago that G\"{o}del's self-referencing results in mathematics
might be relevant to understanding cosmogony.  So, as in  QFT, there is the suggestion of some 
intrinsic randomness.  Of course in the case of arithmetic the randomness arises even in a
sufficiently complex object-based logic system.  We conjecture that in a completely self-referential
system the necessity for SRN is even stronger. 

A further clue to the
fundamental presence of non-local SRN is that of the randomness of the quantum measurement
process, and particularly its non-local manifestations as clearly revealed by 
Bell\cite{Bell}.  We will argue that the peculiarities of the quantum measurement process
are manifestations of the non-linear and non-local character of a HPS via its
objectification process, which proceeds along the lines of a  localising collapse  of 
large-scale non-local configurations, induced by macroscopic objects (detectors in the case of
quantum measurements).  

\section {Heraclitean Process Systems\label{section:Heraclitean-Process-System}}  

To construct a complete bootstrap model of the  universe we must  attempt to take into
account these various considerations and arrive at a HPS showing SOC characteristics with
a fractal 3-space as the dominant emergent {\it universal} feature. The general 
requirements are then randomness, non-linearity, non-locality (this is actually automatic in
 the sense that we must not build in  the notion of  geometry and thus of locality; and so our model must by
definition be non-geometric), and finally iterative in order to generate fractal structures.
 
One  such realisation  is suggested by extending the QFT deconstruction  begun in
\cite{CK}. There we  took the bilocal field representation (hence the origin of the
notation $B$ below) of Quantum Electrodynamic  type field theories and essentially `removed' the
underlying geometrical
 Euclidean space. All these considerations\cite{CK,CK2} suggested the following non-linear
noisy iterative map as our first HPS realisation:

\begin{equation}
B_{ij} \rightarrow B_{ij} -(B+B^{-1})_{ij}\eta + w_{ij},  \mbox{\ \ } i,j=1,2,...,2M; M \rightarrow
\infty.
\label{eq:map}\end{equation} 
In this modelling  we introduce, for convenience only, some terminology: we think of
$B_{ij}$ as indicating the connectivity or relational strength between two monads
$i$ and $j$. The monads concept was introduced by Leibniz, who espoused the {\it relational}
mode of thinking in response to and in contrast to Newton's {\it absolute} space and time. 
Leibniz's ideas were very much in the {\it process} mould of thinking.
It is important to note that the 
 iterations of the map do not constitute {\it a priori} the phenomenon of time, since they are
to perform the function of producing the needed fractal structure.  
 The monad  $i$ acquires
its meaning entirely by means of the  connections
$B_{i1}, B_{i2},...$, where
$B_{ij}=-B_{ji}$  avoids  self-connection ($B_{ii}=0$), and real number valued. The
map in (\ref{eq:map}) has the form of a Wiener process, where the $w_{ij}=-w_{ji}$ are
independent random variables for each $ij$ and for each iteration, and  with variance $\eta$. 
 The $w_{ij}$   model the self-referential noise.  The `beginning' of a universe is
modelled by starting the iterative map with $B_{ij} \approx 0$, representing the absence of
order. Clearly due to the $B^{-1}$ term iterations will rapidly move the
$B_{ij}$ away from such starting conditions.

The non-noise part of the map involves $B$ and
$B^{-1}$. Without the non-linear inverse term the map would produce independent and trivial random
walks for each
$B_{ij}$ - the inverse introduces a linking of all monads. We have chosen $B^{-1}$
because of its indirect connection with quantum field theory\cite{CK}  and
because of its self-organising property. It is the conjunction of the noise and non-noise
terms which leads to the emergence of self-organisation.  Hence the map
models a non-local and noisy relational  system  from which we extract spatial and time-like
behaviour, but we expect residual non-local and random processes characteristic of quantum
phenomena including Einstein-Padolsky-Rosenfeld (EPR)/Aspect type effects. There are  several  other
proposals considering noise in spacetime modelling\cite{Percival,Cal}.

\section  {Emergent Space\label{section:Emergent-Space} }

Here we discuss  this  HPS iterative map, analysis of which suggests the emergence of a 
dynamical 3-dimensional fractal spatial structure. Our results follow from a combination of
analytical and numerical studies.  Under the mapping the noise term will produce rare large
value
$B_{ij}$. Because the order term is generally much smaller, for small
$\eta$, than the disorder term these large valued $B_{ij}$ will persist under the mapping  through more
iterations than smaller valued $B_{ij}$. Hence the larger $B_{ij}$ correspond to some
temporary background structure which we now identify.  

Consider the connectivity  from the point of
view of one monad, call it monad $i$.  Monad $i$ is connected via these large $B_{ij}$ to a
number of other monads, and the whole set of connected monads forms a tree-graph relationship. This is
because  the large links are very improbable, and a tree-graph relationship is much more probable
than a similar graph involving the same monads but with additional links. The set of all large
valued $B_{ij}$ then form tree-graphs disconnected from one-another; see Fig.1a.  In any one
tree-graph the  simplest `distance' measure for any two nodes within a graph is  the smallest number
of links  connecting  them. Indeed this distance measure arises naturally using matrix
multiplications when the connectivity of a graph is encoded in a connectivity or adjacency matrix. 
Let
$D_1, D_2,...,D_L$ be the number of nodes of distance
$1,2,....,L$ from node $i$ (define $D_0=1$ for convenience), where $L$ is the largest distance
from $i$ 
in a particular tree-graph, and let $N$ be the total number of nodes in the tree. Then we have
the constraint $\sum_{k=0}^LD_k=N$. See Fig.1b for an example.

\hspace{-18mm}\begin{minipage}[t]{60mm}
\hspace{15mm}\includegraphics[scale=0.4]{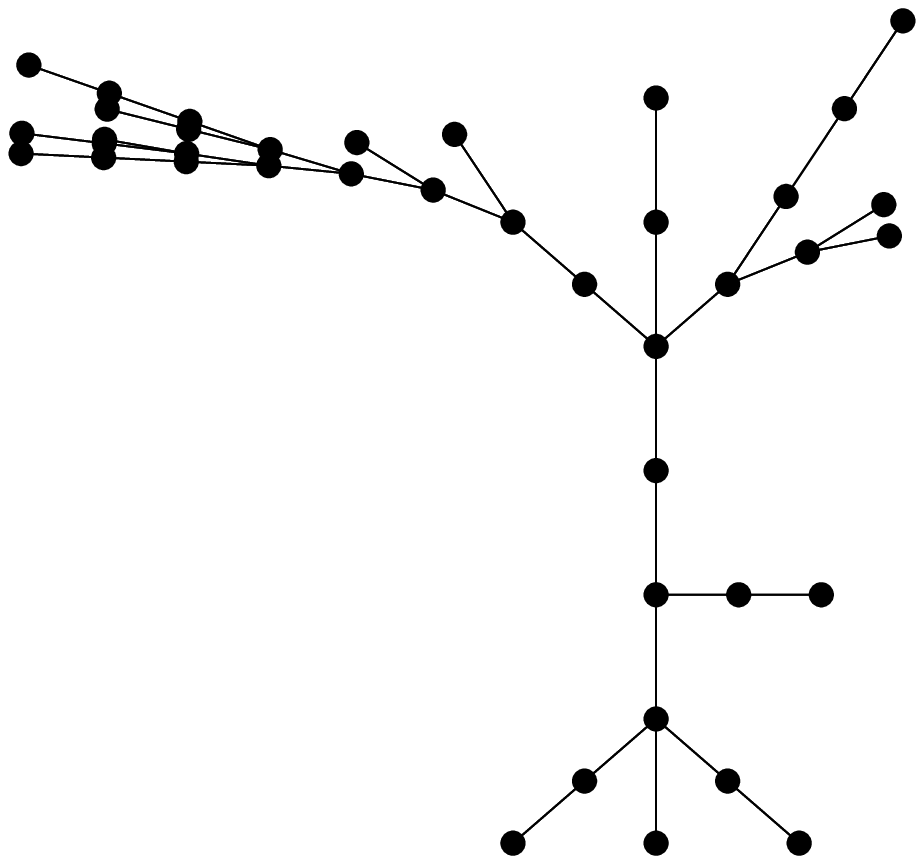}
\makebox[130mm][c]{(a)}
\end{minipage}
\begin{minipage}[t]{60mm}
\hspace{10mm}\includegraphics[scale=0.4]{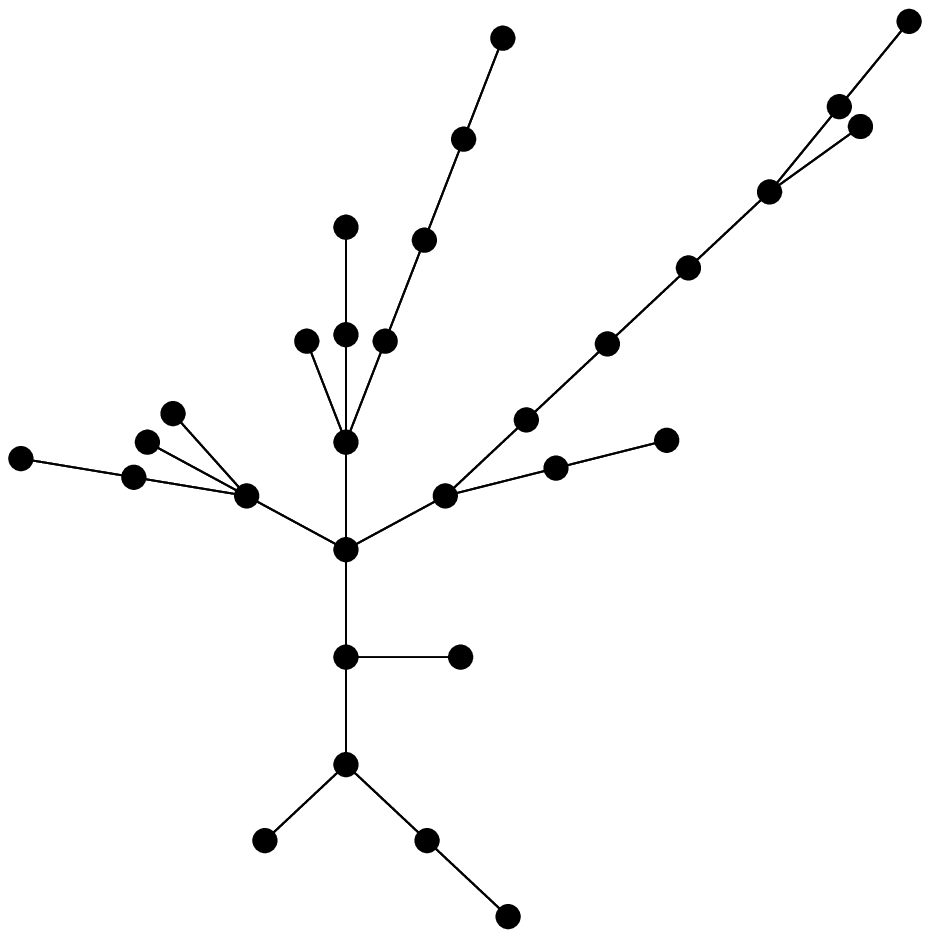}
\end{minipage}
\begin{minipage}[t]{50mm}
\setlength{\unitlength}{0.20mm}

\hspace{3mm}\begin{picture}(0,200)(100,0)  
\thicklines

\put(155,205){\line(3,-5){60}}
\put(155,205){\line(-3,-5){60}}
\put(115,140){\line(3,-5){42}}
\put(195,140){\line(-3,-5){21}}

\put(135,200){ \bf $i$}
\put(225,200){ \bf $D_0\equiv 1$}
\put(225,140){ \bf $D_1=2$}
\put(225,100){ \bf $D_2=4$}
\put(225,65){ \bf $D_3=1$}

\put(155,205){\circle*{5}}

\put(115,140){\circle*{5}}
\put(195,140){\circle*{5}}

\put(95,105){\circle*{5}}
\put(135,105){\circle*{5}}
\put(175,105){\circle*{5}}
\put(215,105){\circle*{5}}

\put(155,70){\circle*{5}}
\end{picture}

\makebox[25mm][c]{(b)}
\end{minipage}
\begin{figure}[ht]
\vspace{-5mm}\caption{\small (a) Rare and large components of $B$ form disconnected
tree-graphs, (b) An $N=8$ tree-graph with $L=3$ for monad {\it i}, with indicated distance
distribution $D_k$.
 \label{figure:gebits}}
\end{figure}

\begin{figure}[ht] 
\hspace{50mm}\includegraphics[scale=0.6, bb=0 20 240 228]{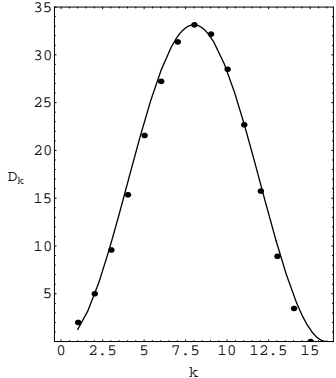}
\caption{\small Data points shows numerical solution of eq.(3) for distance distribution $D_k$ for
a most probable tree-graph  with $L=16$. Curve shows fit of approximate analytic form 
$D_k\sim \mbox{sin}^2(\pi k/L)$ to numerical solution, indicating weak but natural embeddability in
an $S^3$ hypersphere.
 \label{figure:Plot}}
 \end{figure}

 Now consider
the number ${\cal N}(D,N)$ of different $N$-node trees, with the same distance distribution
$\{D_k\}$, to which $i$ can belong.  By counting the different linkage patterns, together with
permutations of the monads we obtain
\begin{equation}
{\cal N}(D,N)=\frac{(M-1)!D_1^{D_2}D_2^{D_3}...D_{L-1}^{D_L}}{(M-N-2)!D_1!D_2!...D_L!},
\label{eq:analytic}\end{equation}
 Here $D_k^{D_{k+1}}$ is the number of different possible linkage patterns between level $k$
and level $k+1$, and $(M-1)!/(M-N-2)!$ is the number of different possible choices for the
monads, with
$i$ fixed. 
 The denominator accounts for those permutations  which have already
been accounted for by the $D_k^{D_{k+1}}$ factors. We compute the most likely tree-graph
structure by  maximising   $\mbox{ln}{\cal N}(D,N)+\mu(\sum_{k=0}^L D_k-N)$ where $\mu$ is a
Lagrange multiplier for the constraint. Using Stirling's approximation for $D_k!$ we obtain
\begin{equation}
D_{k+1}=D_k\mbox{ln}\frac{D_k}{D_{k-1}}-\mu D_k+\frac{1}{2}.
\label{eq:iteration}\end{equation}
which can be solved numerically.  Fig.2  shows a typical  result 
obtained by starting eq.(3) with $D_1=2,D_2=5$ and $\mu=0.9$, and giving  $L=16, N=253$.
Also shown is an approximate analytic solution   $D_k\sim \mbox{sin}^2(\pi k/L)$ found by  
Nagels\cite{Nagels}. 
These results imply  that the most
likely tree-graph structure to which a monad can belong  has a distance
distribution $\{D_k\}$  which indicates that the tree-graph is embeddable in
a 3-dimensional hypersphere, $S^3$. Most importantly  monad $i$ has a 3-dimensional  connectivity to
its neighbours, since $D_k\sim k^2$ for small $\pi k/L$.   We call these
tree-graph $B$-sets {\it gebits} (geometrical bits). However $S^3$  
embeddability of these gebits is a weaker result than demonstrating the necessary emergence of a 3-space,
since extra cross-linking connections would be required   for this to produce a strong embeddability.
But that also appears to be the case, as we now see.

The monads for which the $B_{ij}$ are large thus form  disconnected
 gebits.  These gebits however are in turn linked by smaller and more transient $B_{kl}$,
and so on, until at some low level the remaining $B_{mn}$ are  noise only; that
is they will not survive an iteration. Under iterations of the map this  network 
undergoes growth and decay at all levels, but with the higher levels (larger $\{B_{ij}\}$
gebits) showing most persistence. It is convenient  to relabel the monads so that the current
gebits $g_1,g_2,...$ form matrices  block diagonal   within
$B$, and embedded amongst the smaller and more common noise entries. 

A key dynamical feature is that  most gebit  matrices $g$ have $\mbox{det}(g) =0$, since most tree-graph
connectivity matrices are degenerate. For example in the tree in Fig.1b the $B$ matrix has a
nullspace (spanned by eigenvectors with eigenvalue zero) of dimension two irrespective of the actual
values of the non-zero
$B_{ij}$; for instance the right hand pair  ending at the level $D_2=4$  are identically connected
and this causes two rows (and columns) to be identical up to a multiplicative factor. So the
degeneracy of the gebit matrix is entirely structural. For this graph there is also a second set of three
monads whose connectivities are linearly dependent.  These
$\mbox{det}(g) =0$ gebits  form a   {\it reactive gebits}  subclass of all those gebits generated by the SRN. They
are the building blocks of the self-organising process, and we define their {\it reactive monads} as those appearing
in the nullspaces. Because of the antisymmetry of  $B$ in  this model gebits  with an odd number of monads
automatically have a nullspace of dimension $\geq 1$.
  Monads belonging to the nullspace  form the  reactive or dynamical components of a reactive gebit under the 
mapping  because of the $B^{-1}$ order term: in the absence of the noise $B^{-1}$
would be singular for  reactive  gebits, but in the presence of the noise the
matrix is invertible but with large entries connecting the reactive monads within $B$. 

Numerical
studies show that the  outcome from the iterations  is that the gebits are  seen to interconnect by
forming new links between reactive monads and to do so much more often than they self-link as a
consequence of
links between reactive monads in the same gebit. We also see  monads not currently belonging to
gebits  being linked to reactive monads in existing gebits.  Furthermore the new links, in the main,
join monads located at the periphery of the gebits, i.e these are the most reactive monads of the
gebits.  Of course it is the lack of  appropriate  cross-linkings between these particular monads
that  results in individual gebits  having only a weak 
$S^3$-embeddability.   Hence the new links preserve the
3-dimensional environment of the inner gebits, with the outer reactive monads participating in new
links. Clearly once gebits are sufficiently  linked by  $B^{-1}$  they cease to be
reactive and  slowly  die via the iterative map.  Hence there is a on-going changing population
of reactive gebits that arise from the noise, cross-link, and finally decay. Previous
generations  of active but now  decaying cross-linked  gebits are thus embedded in the structure
formed by the newly emerging  reactive gebits.

 These numerical studies thus reveal gebits competing  in a Darwinian
life-cycle.  However we must next  characterise the global   structure formed by this
transient population of  cross-linking gebits. We suspected that the simplest global structure
 might   correspond to an  emergent geometry in that the dominant links defined a
structure strongly embeddable in a $S^3$ hypersphere, but with the weaker links diffusing the
embedding. To test this hypothesis we first ran the iterative map with a modest $N=100$
monads for some 10 iterations, but with our random SRN term biased to produce a greater number
of large $B_{ij}$.  To test for embeddability we then minimised with respect to the
monad positions  an
$E^n$ embedding measure defined by
\begin{equation}
V(X)=\sum_{i>j}B^2_{ij}\left(D(X^i,X^j)-\frac{1}{|B_{ij}|}\right)^2
\label{eq:embed}\end{equation}
where $X^i=\{x^{(i)}_1,...,x^{(i)}_n\}$ is the possible Euclidean position of monad $i$ 
in $E^n$, and $D(X^i,X^j)=\surd(\sum_{\alpha=1,n}(x^{(i)}_\alpha-x^{(j)}_\alpha)^2)$ is the Euclidean
distance between monad $i$ and monad $j$ in $E^n$. The measure $V$ corresponds to a {\it spring}
embedding model in which the spring between monad $i$ and $j$ has spring constant
$\kappa=B_{ij}^2$, and natural length $1/|B_{ij}|$. The minimisation with respect to the
$X$'s then minimises the `energy' stored in the
$n$-dimensional network of springs in such a manner that strongly linked monads have a strong
spring linking them with a short  natural length, so that the minimisation attempts to
place them at short separation, while  the  weaker monad links (smaller
$|B_{ij}|$) are  represented by weaker and longer  springs, so that these links have less
influence on the embedding, and are allowed to diffuse any $E^n$ embedding.
For the same  $B_{ij}$ matrix we then performed the minimisation of $V$ {\it wrt} the $100n$
monad coordinates for spaces of dimension $n=2,3,4$. We then searched the resulting
embedding for an $S^{n-1}$ signature, noting that $S^{n-1}$ is embeddable in $E^n$ with all
points at the same radius from the `centre'. Hence for each embedding in $E^n$ we located the
`centre-of-mass' (com) of the located monad set, defined by 
$x^{\mbox{com}}_\alpha=\frac{1}{N}\sum_{i=1,N}x^{i}_\alpha$, and computed the radial distance
$R^{(i)}$ of each monad from this  centre according to
$R^{(i)}=\surd(\sum_{\alpha=1,n}(x^{\mbox{com}}_\alpha-x_\alpha^{(i)})^2)$.  We then extracted the
radial distribution   of the monads in each $E^n$, but with the monad $i$ contributing to this
distribution with a weighting   $w^{(i)}$ proportional to the maximum value of $B_{ij}^2$ wrt all
$j$; this ensures that it is the location of the most strongly linked monads which dominate these
distribution plots, since they had the greatest influence on the embedding. 

\begin{figure}[ht] 
\hspace{25mm}\includegraphics[scale=0.8]{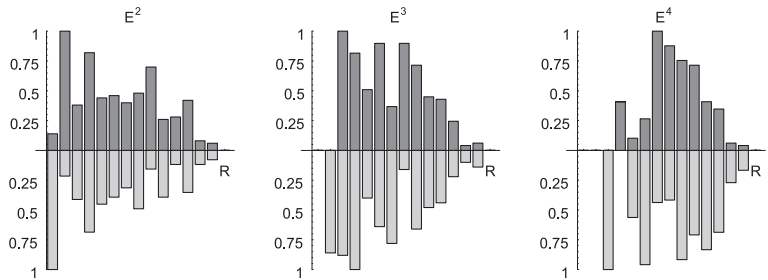}
\caption{\small Above axis plots shows density of monads   plotted against radial distance after
embedding in $E^n$ for $n=2,3,4$. Below axis plots shows randomly generated cases for same
number of monads. A peaking for $E^4$ case indicates a strong embeddability of $B$ in an $S^3$
hypersphere.
 \label{figure:Embed}}
\end{figure}

 The results are shown in
Fig.3  together with the distribution obtained from a purely random embedding:  this in the limit of
large $N$ would produce a flat distribution, but for $N=100$ the statistical variations are
noticeable. The key result of these studies is that the larger valued  entries of the $B$-matrix show the
presence of a
$S^3$ structure as revealed by  the peaking of the radial distribution for  the case of $E^4$.  For $E^2$ and $E^3$
the radial distribution is essentially flat and no different from that generated by a random embedding.   For $E^5$
our analysis would reveal the embeddability already apparent for $E^4$.  The finite width of the peak in the $E^4$
embedding shows  links that are outside of the $S^3$ geometry, and thus are non-local {\it wrt} the $S^3$ geometry.
We conjecture that this effect is an indication of {\it quantum phenomena} attached non-locally to an emergent
geometrical 3-space.

Hence we have produced some evidence  that a three-dimensional geometry is the dominant emergent structure. 
  First we saw that, of the disconnected randomly produced tree-graphs, the most probable  forms have
a three-dimensionality potential; these we called  gebits.
 Second that in turn  most of these  form reactive  gebits  which  cross-link via
the iterative map to produce a strong embeddability in $S^3$  (Of course with only 100 monads any
embeddability will be noisy, and not all runs succeeded  in producing the $S^3$ signature in an $E^4$ embedding,
shown in Fig.2).  After the HPS has produced an $S^3$ structure we have the notion of {\it local} in
the sense of having a {\it position}.  As we have previously discussed, it is required that the
resultant  3-space be fractal in order that the details of the particular HPS realisation are hidden
via
 {\it universality}. Clearly the numerical examples above are inadequate to reveal fractal structure,
though the decay of older gebits does suggest a fractal structure.  A numerical demonstration of 
a fractal 3-space would require a huge increase in monad numbers, or alternatively the development of analytical
techniques.

 Hence
in combination the order and disorder terms synthesise a dynamical 3-space which
 is entirely relational; it does not arise within any {\it a priori} geometrical background
structure.  By  construction it is the most robust structure  - however other softer emergent quantum modes
of behaviour will be seen  attached to   this flickering 3-space.

\section {Emergent Time\label{section:Emergent-Time}}  

It is important to note that the 
 iterations of the map do not constitute {\it a priori} the phenomenon of time, since they perform
the function of producing the needed fractal structure.  However the analysis to  reveal the
internal experiential time phenomenon  is non-trivial, and one would certainly hope to recover the
local nature of experiential time as confirmed by special and general relativity experiments. 
However (experiential) time is only predicted in this model if there is an emergent ordered
sequencing of events at the level of {\it universality}, i.e. above  the details which are purely
incidental to any particular realisation. As a counter example it could well be the case that the
iterative map fails to produce {\it change} at a high level; in which case there is no emergent time
phenomenon.     

 However in the absence of a {\it universality} analysis we take our clues from the
nature of the iterative map  and notice that the modelling of the time phenomenon
here is expected to be much richer than that of the historical/geometric modelling. First the map is
clearly  uni-directional   as there is no way to even define an inverse mapping because of the role
of the noise term, and this is very unlike the conventional time-symmetric differential equations of
traditional physics. In the analysis of the gebits we noted that they show strong persistence, and
in that sense the mapping shows a natural partial-memory phenomenon, but the far `future' detailed
structure of
 even this spatial network is completely unknowable without performing the iterations.
Furthermore the sequencing of the spatial and other structures is individualistic  in that a
re-run of the model will always produce a different outcome. Most important of all is that
we also obtain a modelling of the `present moment' effect, for  the outcome
of the next iteration is   contingent on the noise. So the system shows
overall a sense of a recordable past, an unknowable future and a contingent present moment. 

The HPS process model is  expected to be capable of a better modelling of our experienced
reality, and the key to this is    the noisy processing    the model requires. As well we need
the  `internal view', rather than the `external view' of conventional modelling in physics.
Nevertheless  we would expect that the internally recordable history could be indexed by the
usual real-number/geometrical time coordinate at the level of {\it universality}. 

This new self-referential process modelling requires a new mode of analysis  since one cannot
use externally imposed meta-rules or interpretations, rather, the internal experiential phenomena
and the characterisation of the simpler ones by emergent `laws' of physics must be  carefully
determined.  There has indeed been one  ongoing study  of how  (unspecified) closed
self-referential  noisy  systems acquire self-knowledge and how the emergent
hierarchical structures can `recognise' the same `individuals'$^{\cite{BK}}$.  We believe that our
HQS process model may provide an explicit representation for such studies.

\section  {Objectification\label{section:Objectification}}  

To demonstrate  the viability of the HPS model of reality  we must exhibit an extremely important
but often ignored process, namely {\it objectification}. Despite its many successes the quantum
theory has one significant failure, namely in not predicting the spatial localisation of macroscopic
quantum systems in such a manner  that they behave as familiar {\it objects}.  Schr\"{o}dinger,
Einstein and others were quick to identify this failure, which is a direct consequence of the
linearity of quantum theory.  

The objectification process has largely gone undiscussed simply because
the prevailing attitude is that even  electrons and protons are small objects (particles) which are
localised but whose actual position can only be predicted statistically.  There is in fact no
experimental evidence for this notion.  What is supported by experiment and the quantum theory is 
that the probability of a localisation in space {\it by}  a detector is probabilistic with the
probability given by the usual Born prescription ($\propto|\psi(x)|^2$). The quantum theory itself
makes no reference to objects, and indeed the often quoted {\it radii} of the proton and other quantum states
are nothing more than  correlation lengths.  

In the early stages of constructing Heraclitean Process Systems one
can only outline the expected nature of the emergent quantum systems and of the objectification
process, which we now briefly discuss.
The evidence is that the primary and dominant process is the emergence of a dynamical 3-space,
and that this arises from a non-linear noisy map. From this one would expect secondary 
 phenomena corresponding to small deviations from the localisation  process inherent in the
dominance of the 3-space, that is, the strongly non-linear system will only support small amplitude
non-local  deviations whose description would be by a linear theory:  these we expect to be essentially the well
known quantum phenomena.  And so the HPS is then best described as sub-quantum.  However macroscopic quantum systems
(formed by bound states of large numbers of quantum systems) will amount to large amplitude non-local
deviations from the 3-space process, and the strong non-linearity of the underlying system  would not
support the continued persistence of such a non-local deviation, since this deviation is
incompatible with the formation of the 3-space.  Indeed one would expect the complete system  to
undergo a sudden restoration of  3-space; amounting to a localisation of the macro-quantum system. 
This is the objectification process. Penrose$^{\cite{Penrose}}$  has proposed  a quantum gravity driven
Objective-Reduction  process.  The HPS appears to amount to an  implementation of that suggestion which is,
however, sub-quantum since the HPS apparently generates the spatial phenomena without proceeding through a
quantum description.  Not only would such a mechanism be responsible for the ongoing localisation and consequent
emergence of objects, but should a small amplitude non-local (quantum) deviation result in one or more detectors
(localised objects) becoming non-local  by interacting with that quantum process one would expect the
objectification or localisation process again to be strongly manifested  but to do so randomly via the non-local
SRN.  This means that the localisation process must be non-local in character and not to be understandable as
the propagation of some signal locally through the 3-space.  Such phenomena are seen in  EPR/Aspect experiments
with their apparent concomitant faster-than-light non-local effects.   

The objectification process is also essential to recovering our
everyday object-based phenomena and  its logic. A demonstration of  emergent objectification would then amount to a
derivation of  logic from a   sub-quantum modelling, which itself only used such logic in as much as
it could be suppressed by the SOC argument.    Omn\`{e}s$^{\cite{Omnes}}$  has also discussed the
derivation of logic but from quantum theory via the decoherence mechanism for objectification.
However that mechanism imports  objectification by assuming the quantum measurement postulate in an
{\it ad hoc} manner. 

 There is evidence that the quantum measurement process does indeed involve non-local SRN at the
sub-quantum level, for  it has been discovered that  the  individuality of the
measurement process - the `click' of the detector - can be modelled by adding a noise term to the
Schr\"{o}dinger equation$^{\cite{GP}}$. Then by performing an ensemble average over many individual
runs of this non-linear and stochastic  Schr\"{o}dinger equation one can derive the ensemble
quantum measurement postulate - namely
$<A>=(\psi,A\psi)$ for the ``expectation value of the operator A''.

\section {Conclusion\label{section:Conclusion}}
 
We have addressed here the unique end-game problem which arises when we attempt to model and comprehend
the universe as a closed system without assuming high level phenomena such as space, time and objects
- nor even
 object-based logic. To do this we have proposed a bootstrap modelling which invokes self-organised criticality
to allow the start-up mechanism of the bootstrap to be hidden.    The outcome is the suggestion that
the peculiarities of this end-game problem are directly relevant to our everyday experience of space (and
time); particularly the phenomena of the three-dimensionality of  space (and elsewhere of the contingent present
moment).    This analysis is based upon the  notion that a closed
self-referential system, and  the universe is {\it ipso facto} our only true instance,  is necessarily noisy. This
follows as a conjectured generalisation of the work of G\"{o}del and Chaitin on self-referencing in
the abstract and artificial game  of mathematics.  To explore the implications we have considered a simple {\it
pregeometric  non-linear noisy iterative map}.  The analysis of this map shows that the first self-organised
structure to arise  is a dynamical 3-space formed from competing pieces of 3-geometry - the gebits; however the
actual details of this level of modelling are necessarily to be hidden via the self-organised
criticality of the model.  The analysis of experiential time is more difficult, but it will clearly
be a contingent and process  phenomenon which is more complex and hence richer than the current
geometric/historic modelling of time.  We suggest that the non-local
self-referential noise has been a  major missing component of traditional modelling of reality.  

We acknowledge useful discussions with S.M. Gunner and K. Kitto.
Research supported by an ARC Small Grant from Flinders University.

\end{document}